\documentclass[twocolumn,prd,showpacs,superscriptaddress,preprintnumbers,nofootinbib]{revtex4}
\usepackage{graphicx,epsfig}
\usepackage{bm}
\usepackage{latexsym,amssymb,amsmath,float}
\usepackage{color}

\newcommand{\beq}[1]{\begin{equation}\label{#1}}
\newcommand{\eeq}{\end{equation}}
\newcommand{\bea}[1]{\begin{eqnarray} \label{#1}}
\newcommand{\eea}{\end{eqnarray}}
\newcommand{\ba}{\begin{array}}
\newcommand{\ea}{\end{array}}
\newcommand{\nn}{\nonumber}
\newcommand{\rf}[1]{(\ref{#1})}
\def\be{\begin{equation}}
\def\ee{\end{equation}}
\def\gs{\mathrel{
   \rlap{\raise 0.511ex \hbox{$>$}}{\lower 0.511ex \hbox{$\sim$}}}}
\def\ls{\mathrel{
   \rlap{\raise 0.511ex \hbox{$<$}}{\lower 0.511ex \hbox{$\sim$}}}}

\newcommand{\eps}{\epsilon}

\newcommand{\rarr}{\rightarrow}

\newcommand{\nue}{\nu_{e}}
\newcommand{\numu}{\nu_{\mu}}

\newcommand{\numubar}{{\bar \nu}_\mu}

\newcommand{\epspi}{\epsilon_{\pi^\pm}}
\newcommand{\postscript}[2]{\setlength{\epsfxsize}{#2\hsize}
   \centerline{\epsfbox{#1}}}

\begin{document}

\title{Pinning down the cosmic ray source mechanism with new IceCube data}

\author{Luis A.~Anchordoqui}
\affiliation{Department of Physics,
University of Wisconsin-Milwaukee,
 Milwaukee, WI 53201, USA
}

\author{Haim Goldberg}
\affiliation{Department of Physics,
Northeastern University, Boston, MA 02115, USA
}

\author{Morgan \nolinebreak H. \nolinebreak Lynch}
\affiliation{Department of Physics,
University of Wisconsin-Milwaukee,
 Milwaukee, WI 53201, USA
}

\author{Angela V. Olinto}
\affiliation{Department of Astronomy and Astrophysics, Enrico Fermi
Institute, University of Chicago, Chicago, Il 60637, USA}
\affiliation{Kavli Institute for Cosmological Physics,  University of Chicago, Chicago, Il 60637, USA}

\author{Thomas C. Paul}
\affiliation{Department of Physics,
University of Wisconsin-Milwaukee,
 Milwaukee, WI 53201, USA
}

\affiliation{Department of Physics,
Northeastern University, Boston, MA 02115, USA
}

\author{Thomas J. Weiler}
\affiliation{Department of Physics and Astronomy,
Vanderbilt University, Nashville TN 37235, USA
}

\date{June 20, 2013}
\begin{abstract}
  \noindent
  Very recently the IceCube Collaboration has reported an observation
  of 28 neutrino candidates with energies between $50$~TeV and
  $2$~PeV, constituting a 4.1$\sigma$ excess compared to the
  atmospheric background.  In this article we investigate the
  compatibility between the data and a hypothesized unbroken power-law
  neutrino spectrum for various values of spectral index $\Gamma \ge
  2$.  We show that $\Gamma \sim 2.3$ is consistent at the $\sim
  1.5\sigma$ level with the observed events up to 2~PeV and to the
  null observation of events at higher energies.  We then assume that
  the sources of this unbroken spectrum are Galactic, and deduce {\it
    (i)} an energy-transfer fraction from parent protons to pions
  (finding $\epspi$ and $\eps_\pi$), and {\it (ii)} a way of
  discriminating among models which have been put forth to explain the
  ``knee'' and ``ankle'' features of the cosmic ray spectrum.  Future
  IceCube data will test the unbroken power law hypothesis and provide
  a multi-messenger approach to explaining features of the cosmic ray
  spectrum, including the transition from Galactic to extragalactic
  dominance.

\end{abstract}


\maketitle

\section{Introduction}

In April 2013 the IceCube Collaboration published an observation of
two $\sim$~1 PeV neutrinos, with a p-value 2.8$\sigma$ beyond the
hypothesis that these events were atmospherically
generated~\cite{Aartsen:2013bka}.  These two candidates were found in
a search for events with a significant energy deposition as expected
for cosmogenic neutrinos~\cite{Beresinsky:1969qj}. Results of a new
search technique designed to extend the range of energy sensitivity
were reported in~\cite{Aartsen:2013jdh}.  In the new search protocol,
selected events were required to start inside an inner fiducial volume
of the detector. The fact that neutrinos are produced in high energy
cosmic ray events means that atmospheric neutrinos of sufficiently
high energy and sufficiently small zenith angle will be accompanied by
a muon of the same event and therefore excluded from the sample as
entering muons. The veto has been derived explicitly
in~\cite{Schonert:2008is} only for muon neutrinos accompanied by the
muon from the same decay.  This technique is particularly effective
for energies $E_\nu > 100$~TeV and zenith angles less than $60^\circ$
or $70^\circ$, where the boost is sufficient to ensure that the shower
muons and neutrinos follow nearly identical trajectories.

The new analysis revealed an additional 26 neutrino
candidates depositing ``electromagnetic equivalent energies'' ranging
from about 30~TeV up to 250~TeV.  Seven of the events show visible
evidence of a muon track, and the remainder are consistent with
cascade events. The quoted background estimate from atmospheric
neutrinos is $10.6^{+5.0}_{-3.6}$. Taken together, the total sample of
28 events departs from the atmospherically-generated neutrino
hypothesis by $4.1\sigma$.

Interpreting these results in terms of popular astrophysical models
appears to be challenging.  First of all, if the neutrino flux is
indeed a Fermi-shock flux falling as an unbroken $E_\nu^{-2}$
power-law, one would expect about 8-9 events above 1~PeV, which thus
far are not observed.  This null result at high-energy may be
indicative of a cutoff in the spectrum at
$1.6^{+1.5}_{-0.4}$~PeV~\cite{Aartsen:2013jdh}.  On the other hand, the null
result may indicate a steeper but still unbroken $E_\nu^{-\Gamma}$
spectrum, with $\Gamma > 2$.  An interesting issue for the future is
whether the data offers directional information about the sources.
Neither auto-correlation studies of the data, nor cross-correlation
studies of the data with candidate source types, have yet been
reported.  The IceCube angular resolution for shower events is poor,
$15^\circ$, so firm conclusions are elusive at present.

In this work we investigate the compatibility between the IceCube
observations and the hypothesis of an unbroken power-law spectrum
arising from optically thin Galactic neutrino sources. The layout of
the paper is as follows. We begin in Sec~\ref{s-dos} by studying which
are the possible source spectral indices that are consistent with the
data reported thus far. We next assume that the neutrino sources are
Galactic in origin, and turn our attention to two interesting
consequences of the Galactic power-law hypothesis.  The first is the
implication for spectral features observed in the cosmic ray (CR)
energy spectrum.  The second is an implication for the average
efficiency of the energy transfer from protons to the charged pions
which decay to yield the cosmic neutrino flux. Namely, assuming the
neutrinos are indeed of Galactic origin, in Sec.~\ref{s-tres} we
explore what IceCube data may tell us about competing theories
describing the region of the (baryonic) cosmic ray flux transition
from Galactic to extra-galactic dominance.  After that, in
Sec.~\ref{s-cuatro} we deduce the energy transfer fraction from the
parent protons to the pions which ultimately produce the observed
neutrinos, demonstrating that $pp$ collisions are more likely to
produce the neutrino flux than are $p \gamma$ collisions. As we based
our arguments on the hypothesis that the IceCube neutrino excess
originates from optically thin Galactic sources, it is natural to
consider evidence which may corroborate or refute this hypothesis.
Very recently it has been argued that existing photon bounds could
call into question the possibility of a (predominantly) Galactic
origin for the IceCube neutrino excess~\cite{Ahlers:2013xia}.  In
Sec.~\ref{s-cinco} we revisit the subject, considering the impact of
photon bounds as well as other factors relevant to the veracity of our
model.  Finally in Sec.~\ref{s-seis} we make a few observations on the
consequences of the overall picture discussed herein.

In a complementary fashion to this paper, other authors have recently
explored potential extragalactic neutrino sources~\cite{Cholis:2012kq}
and new massive particle physics~\cite{Feldstein:2013kka} as
explanations of the IceCube data. For a recent review
see~\cite{Anchordoqui:2013dnh}.

\section{Spectral Shape}
\label{s-dos}

Herein we hypothesize that the  cosmic neutrino
flux per flavor,  averaged over all three flavors,
follows an unbroken power law of the form
\begin{equation}
\frac{dF_\nu}{d\Omega dA dt dE_\nu} = \Phi_0 \ \left(\frac{E_\nu}{1~{\rm GeV}} \right)^{-\Gamma}\,,
\label{eqn:flux}
\end{equation}
for a factor of several or more above the highest energies so far
observed.  We ask ``What value(s) of the spectral index $\Gamma$ (if
any) are consistent with the recent IceCube observations?''  We
partition observations into three bins: {\it (i)} 26 events from
50~TeV to 1~PeV, which includes the $\sim 10$ atmospheric background
events; {\it (ii)} 2 events from 1~PeV to 2~PeV; {\it (iii)} zero
events above 2~PeV, say from 2~PeV to 10~PeV, with a background of
zero events.

\begin{figure}[tbp]
\postscript{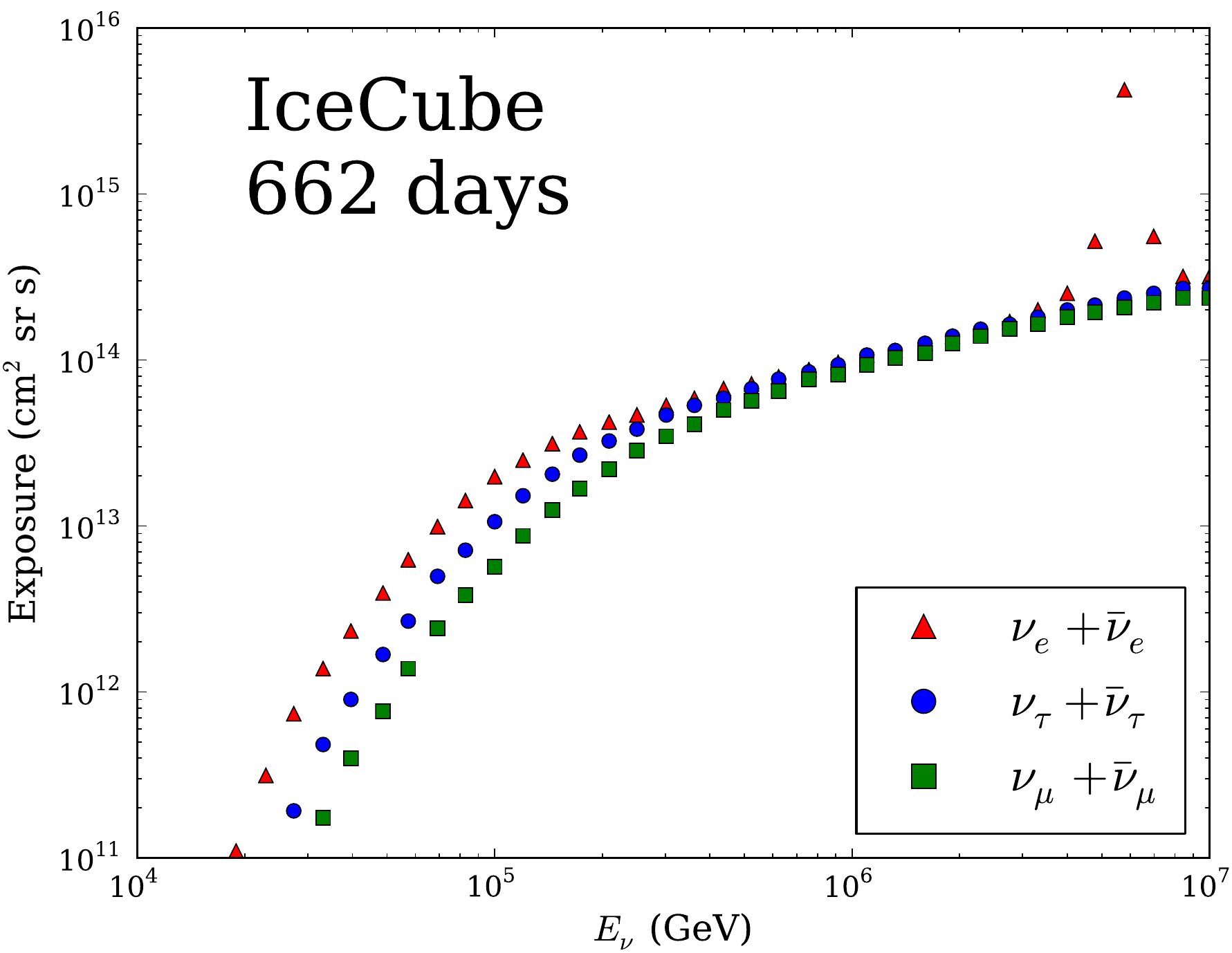}{0.9}
\caption{IceCube exposure for 662 days of data collection. The sharp-peaked structure for $\overline \nu_e$ at 6.3~PeV is due to the Glashow resonance. }
\label{fig:exposure}
\end{figure}

For various spectral indices from 2.0 to 2.8, we fit the neutrino flux
to each of these three bins, by integrating over the energy span of
the bin.  A key point is that we employ IceCube's energy-dependent,
flavor-dependent exposure functions for the 662 days of observation
time reported thus far.  The IceCube exposures are shown in
Fig.~\ref{fig:exposure}.

\begin{table}[tb]
\caption{Flavor-averaged normalization $\Phi_0$ for the ``low energy'' ($E<1$~PeV) and ``high energy'' (1-2~PeV) bins ,
and normalization upper limits for the ``null'' bin (2-10~PeV) at 68\%CL ($\Phi^{\rm max}_{68}$)
and 90\%CL ($\Phi^{\rm max}_{90}$) in units of $({\rm GeV}\cdot{\rm cm}^{2}\cdot{\rm s}\cdot{\rm sr})^{-1}$,
for various spectral indices, $\Gamma$.
\label{tab:cls}}
\centering
\begin{tabular}{ |c|c|c|c|c|}
\hline \hline \\ [-2.0 ex]
$\Gamma$~&~$\Phi_0^{E_\nu < 1{\rm PeV}}$~&$\Phi_0^{1 {\rm PeV} < E_\nu < 2
  {\rm PeV}}$  &~$\Phi^{\rm max}_{68}$~&~$\Phi^{\rm max}_{90}$~\\
\hline \\ [-2.0 ex]
~~2.0~~ & $1.66 \times 10^{-8}$ & $9.50 \times 10^{-9}$ & $3.94 \times
10^{-9}$ & $7.44 \times 10^{-9}$ \\
\hline \\ [-2.0 ex]
2.1 & $5.70 \times 10^{-8}$ & $3.91 \times 10^{-8}$ & $1.84 \times
10^{-8}$ & $3.49 \times 10^{-8}$ \\
\hline \\ [-2.0 ex]
2.2 & $1.95 \times 10^{-7}$ & $1.61 \times 10^{-7}$ & $8.62 \times
10^{-8}$ & $1.63 \times 10^{-7}$ \\
\hline \\ [-2.0 ex]
2.3 & $6.63 \times 10^{-7}$ & $6.62 \times 10^{-7}$ & $4.02 \times
10^{-7}$ & $7.61 \times 10^{-7}$ \\
\hline \\ [-2.0 ex]
2.4  & $2.24 \times 10^{-6}$ & $2.72 \times 10^{-6}$  & $1.88 \times
10^{-6}$ & $3.55 \times 10^{-6}$\\
\hline \\ [-2.0 ex]
2.5 & $7.54 \times 10^{-6}$ & $1.12 \times 10^{-5}$ & $8.73 \times 10^{-6}$ & $1.65 \times$ $10^{-5}$ \\
\hline \\ [-2.0 ex]
2.6 & $2.52 \times 10^{-5}$ & $4.59 \times 10^{-5}$ & $4.06 \times
10^{-5}$ & $7.68 \times 10^{-5}$\\
\hline \\ [-2.0 ex]
2.7 & $8.39 \times 10^{-5}$ & $1.88 \times 10^{-4}$ & $1.88 \times 10^{-4}$ & $3.56 \times 10^{-4}$\\
\hline \\ [-2.0 ex]
2.8 & $2.78 \times 10^{-4}$ & $7.71 \times 10^{-4}$& $8.73\times 10^{-4}$  & $1.65 \times 10^{-3}$ \\
\hline
\hline
\end{tabular}
\end{table}

Our results are summarized in Table~\ref{tab:cls}.  Column two (three)
shows the fitted flux normalization $\Phi_0$ for the first (second)
bin.  The null, third bin requires more explanation: According to the
statistics of small numbers~\cite{Feldman:1997qc}, any flux yielding
more than 1.29 (2.44) events in the null 2-10~PeV range of bin three,
is excluded at 68\% CL (90\% CL).  Accordingly, columns four and five
show the maximum flux normalizations allowed by the null bin three, at
the 68\% and 90\%~CL's.

Under the assumption of a single power-law across the three energy
bins, consistency requires that the maximum flux normalization
determined by bin three must exceed the flux normalizations from bins
one and two.  Moreover, the fitted normalizations from bins one and
two should be the same, or nearly so. In terms of the Table columns,
if flux numbers from columns two or three exceed the maximums of
columns 4 and 5, then the fit is ruled out at 68\% and 90\%
CL. Table~\ref{tab:cls} reveals that spectral indices shallower than
2.3 are inconsistent with the data at 90\% CL or more, while indices
shallower than 2.7 are inconsistent at 68\% CL. Only for $\Gamma =
2.3$ are the normalizations from bins one ($E_\nu < 1$~PeV) and two
($1~{\rm PeV} < E_\nu < 2~{\rm PeV}$) quite consistent with each
other, and therefore with an unbroken power law.  The overall
consistency of the $\Gamma=2.3$ power law across all three bins is at
roughly the $1.5\sigma$ level.  We therefore choose $\Gamma = 2.3$ as
our reference value for the unbroken power law hypothesis.  Taking into
account the errors on the background in the first bin reported by
the IceCube collaboration, we find $\Gamma = 2.3 \pm 0.2$ (with
normalization given in the third column of Table~\ref{tab:cls}).
A recent analysis performed prior to the announcement of the 26 events below
1~PeV is consistent with our finding~\cite{Laha:2013lka}.

\section{Transition from Galactic to extragalactic sources}
\label{s-tres}

Above about 10~GeV, the CR energy spectrum is observed to
fall roughly as a power law; the flux decreases nearly three orders of
magnitude per energy decade until eventually suffering a strong
suppression near 60~EeV~\cite{Abbasi:2007sv}. Close examination
reveals several other spectral features. A steepening of the spectrum
from $J(E) \propto E^{-2.67\pm 0.07}$ to $E^{-3.07 \pm 0.11}$ has been
dubbed the ``knee'' occurring at $E_{\rm knee} \approx 3~{\rm
  PeV}$~\cite{Hoerandel:2002yg}.  A less prominent ``second knee'',
corresponding to a further softening $J (E) \propto E^{-3.52 \pm
  0.19}$ appears above 0.3~EeV~\cite{AbuZayyad:2000ay}. At $E_{\rm
  ankle} \approx 3~{\rm EeV}$ a pronounced hardening of the spectrum
becomes evident, generating the so-called ``ankle'' feature~\cite{Bluemer:2009zf}.

The small variations of the spectral index can be interpreted either
as a transition between CR populations or as an imprint of CR
propagation effects. One model posits that {\it extragalactic} protons
dominate the CR composition at and above the second knee, and that the
ankle feature is carved into the spectrum as a result of $e^+
e^-$~pair production when CR protons interact with the cosmic
microwave background (CMB) photons~\cite{Berezinsky:2002nc}.  This model is
often referred to as the ``dip'' model.  In contrast, a second model
proposes that the ankle feature represents a crossover of the two
fluxes, Galactic and extragalactic, with different spectral indices,
signifying a transition from heavy nuclei of Galactic origin to proton
dominance of the extra-galactic spectrum~\cite{Hill:1983mk}.

If the ankle marks the Galactic to extra-galactic CR transition, then
Galactic sources must be able to accelerate nuclei up to about 1-3
EeV~\cite{Anchordoqui:2003vc}.  Assuming that the highest energies
attainable from a Galactic sources scale as $E = Z E_p$, then
protons should be accelerated to only about $1/26$ of the energy of
the ankle, or $E_p \sim 120~{\rm PeV}$.  Proton interactions with
either photons or other (low energy) protons at the acceleration sites
ultimately give rise to neutrinos which carry on average $\sim 1/16$
of the initial proton energy~\cite{Ahlers:2005sn}.  The neutrino
spectrum has the same spectral index as the hard protons at the
source.  Thus, if the proton spectrum follows an unbroken power law up
to a maximum energy of $\sim 120~{\rm PeV}$,  neutrinos produced
by proton interactions in Galactic sources should exhibit an unbroken
power law which extends to roughly 8-10~PeV, but not beyond. On the
other hand, the dip model places the Galactic to extragalactic
transition in the region of the second knee, $E \sim 500$~PeV.  This
implies, by our previous arguments, a maximum Galactic proton energy
26 times smaller, $E_p \sim 20$~PeV, and a maximum neutrino energy
16~times smaller again, $E_\nu \sim 1$~PeV.

So far, no events from 2-10~PeV have been observed.  As we have shown above,
this null result presents only a $1.5\sigma$ downward fluctuation for an
unbroken power law with spectral index $\Gamma=2.3$.  So the jury is out,
awaiting further IceCube data for the 2-10~PeV region.  If an unbroken neutrino
power law is ultimately confirmed all the way to $\sim 10~{\rm PeV}$, this would
naturally favor the ankle transition model, as some fine tuning would then be
required for the dip model.  On the other hand, if future observations continue
the null view of the 2-10~PeV region, then the dip model becomes favored.

It could, of course, be the case that an extragalactic component
contributes beyond $\sim 2$~PeV, although extraordinary fine tuning
would be required for the spectral indices to be the same above and
below the galactic to extragalactic transition. If the Galactic
sources begin to reach the end of their acceleration potential, one
would expect a break in the index, characterized by a steepening of
the spectrum.  In contrast, if an extragalactic contribution with a
shallow spectrum induces the break, a hardening of the spectrum is
expected above the transition.  Ultimately, IceCube will achieve the
capacity to isolate the sources or source populations of the highest
energy neutrinos, delivering the final verdict.

Given that the CR spectrum exhibits breaks at the knee and second
knee, we should ask whether it is plausible for the proton injection
spectrum to be characterized by a single index over the energy range
of interest. If neutrinos are produced at the same sites as the CRs,
then there are two categories of models which may explain these
breaks; the knee may signify the acceleration endpoint of one of two
types of sources~\cite{Biermann:1995qy}, or the knee may result from
magnetic-dependent leakage of particles from the
Galaxy~\cite{Candia:2003dk}. If the latter is correct then the
injected proton spectrum, and hence neutrino spectrum, should follow
an unbroken power law over the energy ranges under discussion
here. Thus, the shape of the neutrino spectrum arriving from the
Galactic disk will also help to discriminate among these competing
knee models.  It may also be the case that neutrino production during
propagation is relevant. Qualitatively speaking, if this effect is
non-negligible but not dominant, one would expect a hardening of the
neutrino spectrum with energy.  In contrast, if neutrinos are
predominantly generated during propagation, the spectrum should soften
with energy~\cite{oai:arXiv.org:astro-ph/0306632}.

To quantify the spectral features characteristic of these two models we
adopt the ``leaky box'' picture, in which CRs propagate freely in the
Galaxy, contained by the magnetic field but with some probability to
escape which is constant in time. The local energy density is given by
\begin{equation}
n_{\rm CR} (E) \equiv \frac{4\pi}{c} J (E) \, \approx Q(E) \
\tau(E/Z),
\label{leaky}
\end{equation}
where $Q(E) \propto E^{-\alpha}$ is the generation rate of primary CRs
and $\tau (E/Z) \propto E^{-\delta}$ is the rigidity-dependent
confinement time~\cite{Gaisser:1990vg}. Fits to the energy dependence
of secondary to primary ratios yield
$\delta=0.6$~\cite{Swordy:1993dz}. For a source index $\alpha \simeq
2.07$, which is close to the prediction of Fermi shock acceleration,
inclusion of propagation effects reproduces the observed
spectrum. However, $\delta = 0.6$ results in an excessively large
anisotropy which is inconsistent with
observations~\cite{Blasi:2011fm}. Consistency with anisotropy can be
achieved by adopting a Kolmogorov index, $\delta =
1/3$~\cite{Biermann:1995qy,Candia:2003dk}. The apparent conflict with
the secondary to primary composition analyses can be alleviated
through small variations of the energy dependence of the spallation
cross sections, or variation in the matter distribution in the
Galaxy~\cite{Biermann:1995qy}.  This hypothesis implies a steeper
source spectrum, $\alpha \simeq 2.34$, which agrees remarkably well
with the fit of an unbroken power law to IceCube data, as discussed
herein.

We consider a model in which cosmic ray leakage is
dominated by Kolmogorov diffusion, $\tau \propto (E/Z)^{-1/3}$, for $E
< Z E_{\rm knee}$, with increasing leakage due to decreasing trapping
efficiency with rising energy, $\tau \propto (E/Z)^{-1}$ for $E \gg Z
E_{\rm knee}$~\cite{Candia:2003dk}. The knee is etched into the spectrum by a transition
from diffusion to drift motion, while the second knee results from a subsequent
transition to quasirectilinear motion.  Each CR nucleus is
affected by drifts at $E \simeq Z E_{\rm knee}$, resulting in a
progressive steepening of the CR spectrum. Since the lighter component
are strongly suppressed above 0.1~EeV we are left with an iron
dominated spectrum which progressively steepens until the overall
spectrum becomes $J(E) \propto E^{-2.67 - 2/3}$, in agreement with
observation of the second knee~\cite{AbuZayyad:2000ay}.

 It is helpful to envision the CR engines as machines
  where protons are accelerated and (possibly) permanently confined by
  the magnetic fields of the acceleration region. The production of
  neutrons and pions and subsequent decay produces neutrinos,
  $\gamma$-rays, and CRs. If the neutrino-emitting source also
  produces high energy CRs, then pion production must be the principal
  agent for the high energy cutoff on the proton spectrum.
  Conversely, since the protons must undergo sufficient acceleration,
  inelastic pion production needs to be small below the cutoff energy;
  consequently, the plasma must be optically thin. Since the
  interaction time for protons is greatly increased over that of
  neutrons due to magnetic confinement, the neutrons escape before
  interacting, and on decay give rise to the observed CR flux. The
  foregoing can be summarized as three conditions on the
  characteristic nucleon interaction time scale $\tau_{\rm int}$; the
  neutron decay lifetime $\tau_n$; the characteristic cycle time of
  confinement $\tau_{\rm cycle}$; and the total proton confinement
  time $\tau_{\rm conf}$: $(i)\; \tau_{\rm int}\gg \tau_{\rm cycle}$;
  $(ii)\; \tau_n > \tau_{\rm cycle}$; $(iii)\; \tau_{\rm int}\ll
  \tau_{\rm conf}$. The first condition ensures that the protons
  attain sufficient energy.  Conditions $(i)$ and $(ii)$ allow the
  neutrons to escape the source before decaying. Condition $(iii)$
  permits sufficient interaction to produce neutrons and
  neutrinos. These three conditions together define an optically thin
  source.  In what follows we assume these three conditions hold for
  some neutrino-emitting sources  in the Galaxy.

As an illustration, we mention astrophysical environments where the
conditions discussed above could hold. The Galactic Center, for
instance, has been proposed as a source
candidate~\cite{Neronov:2013lza,Razzaque:2013uoa}.  These conditions
can also apply in the jets of powerful microquasars where protons can
be efficiently accelerated beyond the knee feature.  Neutrino
production in $p\gamma$~\cite{Levinson:2001as} and
$pp$~\cite{Reynoso:2008nk} collisions has been suggested as a possible
source of neutrinos.  Attaining the maximum observed neutrino energies
for such scenarios may require fine tuning and pushing parameters to
their extrema, to which one may object. The assumption that sources
with the requisite properties do exist is, however, consistent with a
very general estimate of Galactic cosmic-ray power required to match the
observed spectrum up to about the second knee, as well as a rough
estimate of the ratio of heavy nuclei to protons as measured by
KASCADE near the end of the presumed Galactic spectrum. We elaborate upon these points in the next section. Whatever
point of view one may find most convincing, however, we should rely on
future experimental results rather than ``naturalness'' to settle the question.

\section{Power for Galactic Cosmic Rays}
\label{s-cuatro}

Next we turn to the question of what the Galactic power-law model
developed above would imply regarding the average efficiency of
transferring proton energy to charged pions.  Assume that the source
spectral index of  CRs in the range 0.1 - 100~PeV is $\Gamma$ from here on.  Then, following~\cite{Ahlers:2005sn}, we define
the two constants
\begin{equation*}
C_{\rm CR}^p(\Gamma) \equiv \frac{dF_{\rm CR}^p}{dE\,dA\,dt}\,E^\Gamma\,,
\ \
{\rm and\ \ }
C_{\nu}(\Gamma)  \equiv  \frac{dF_{\nu}}{dE\,dA\,dt}\,E^\Gamma\,,
\end{equation*}
where $C_\nu = 4\pi \Phi_0^{\rm total} \, {\rm GeV}^\Gamma$ and
$\Phi_0^{\rm total} = 3 \Phi_0$, given our assumption of flavor
equilibration.  In conventional notation, we next define $\epspi$ to
be the ratio of CR power (energy/time) emitted in charged pions to
that in the parent nucleons.  We also need $\epsilon_\nu$, defined as
the fractional energy in neutrinos per single charged pion decay.  If
the pion decay chain is complete
($\pi^\pm\rarr e+\nue+\numu+\numubar$), then $\epsilon_\nu \simeq 3/4$,
whereas if the pion decay chain is terminated in the source region by energy loss
of the relatively long-lived muon, then $\epsilon_\nu \simeq 1/4$.
Comparing the energy produced in charged pions at the source to the
neutrino energy detected at Earth, one gets the energy conservation
relation
\begin{equation}
\epsilon_\nu\,\epspi \int_{E_1}^{E_2} \frac{dF_{\rm CR}^p}{dE\,dA\,dt}\;E dE =
	\int_{E_{\nu 1}}^{E_{\nu
            2}}\,\frac{dF_{\nu}}{dE_\nu\,dA\,dt}\;E_\nu dE_\nu\,, \nonumber
\end{equation}
where $E_{\nu 1}=\frac{{E_1}}{16}$, and $E_{\nu 2}=\frac{{E_2}}{16}$;
these integrals may be done analytically to yield (for $\Gamma\neq 2$)
$$
\epsilon_\nu \; \epspi \; C_{\rm CR}^p   \; \frac{E_1^{2-\Gamma} -
  E_2^{2-\Gamma}}{\Gamma-2} =  \frac{\left(\frac{E_1\,}{16}
  \right)^{2-\Gamma} -
  \left(\frac{E_2\,}{16}\right)^{2-\Gamma}}{\Gamma-2} \; C_\nu \, .
$$
Then, solving for $\epspi$ we arrive at
\beq{epspi}
\epspi = \left(\frac{1}{16} \right)^{2-\Gamma} \,
\frac{C_\nu(\Gamma)}{\epsilon_\nu\,C_{\rm CR}^p(\Gamma)} \,.
\eeq
The numerology for $C_\nu$ is given in Table~\ref{tab:cls}.
For the favored spectral index $\Gamma=2.3$, we have
\beq{Cnu2.3}
C_\nu(2.3) = 12 \pi\times 6.6\times 10^{-7} \,{\rm GeV}^{2.3}\,({\rm GeV\, s\, cm}^2)^{-1}\,.
\eeq
The constant $C_{\rm CR}^p(2.3)$ is related to the
injection power  of CR protons, $d\eps_{\rm CR}^p/dt$, as follows:
\bea{Ccr2eps}
\frac{d\eps_{\rm CR}^p}{dt}[E_1, E_2] &=& A \int_{E_1}^{E_2} \frac{dF_{\rm CR}^p}{dE\,dA\,dt}\;E\,dE \nn\\
 &=& A\int_{E_1}^{E_2} \left( \frac{dF_{\rm CR}^p}{dE\,dA\,dt}\;E^{\Gamma}\right) E^{(1-\Gamma)} dE  \nn\\
 &=& A\,C_{\rm CR}^p \,\frac{ \left( {E_1}^{(2-\Gamma)}-{E_2}^{(2-\Gamma)}\right)}{\Gamma-2}\,,
\eea
where $A = 4 \pi r^2$ is an appropriately weighted surface area for the
arriving cosmic-ray or neutrino flux.
In~\cite{Gaisser:1994yf}, $A$ is set equal to $4 \pi R_{\rm G}^2\equiv A_0$, where $R_{\rm G}$ is the
Galactic radius,~$\approx 10~{\rm kpc}$.
However, keeping in mind that $\langle r^{-2} \rangle$ diverges as $\ln(R_{\rm G}/2r_{\min})$,
with $r_{\min}$ being the distance to the nearest source,
$A^{-1}$ can easily be a factor of 2 larger than $A_0^{-1}$.
Two independent arguments support such an enhancement.
The first is to simply note that a local void radius of 0.7~kpc gives $A_0/A = 2$.
The second is to note that the thin-disk approximation breaks down at a small distance
$z$ of order of the disk height, leading to a similar guesstimate of integration cutoff
and resulting enhancement factor. Inverting (\ref{Ccr2eps}) and using the fact that
${E_2}^{(2-\Gamma)} \ll {E_1}^{(2-\Gamma)}$, we get the conversion
\beq{eps2Ccr}
C_{\rm CR}^p =  \frac{d\eps_{\rm CR}^p}{dt} [E_1,E_2] \ \frac{(\Gamma-2)\,E_1^{(\Gamma-2)}}{A}\  .
\eeq
How, and how well, is $d\eps_{\rm CR}^p/dt$ known?  The assumption
underlying the leaky box model is that  the energy density
in CRs observed locally is typical of other regions of the
Galactic disk. If so, the total power required to maintain the cosmic
radiation in equilibrium can be obtained by integrating the generation
rate of primary CRs over energy and space. Using (\ref{leaky}), we obtain
\begin{equation}
\frac{d\epsilon_{\rm CR}}{dt} = \int d^3x \int Q(E) \: dE = V_G \frac{4\pi}{c}
\int \frac{J(E)}{ \tau(E/Z)}  dE \, ,
\end{equation}
where $V_G \sim 10^{67}~{\rm cm}^3$ is the Galactic disk
volume~\cite{Gaisser:2005tu}. For $E_{\rm knee} < E < E_{\rm ankle}$,
we conservatively assume that the trapping time in the Galaxy scales
with energy as $\tau = 2 \times 10^{7} (E_{{\rm GeV}}/Z)^{-1/3}~{\rm
  yr}$~\cite{Gaisser:2006xs}. (Note that an evolution into
quasirectilinear motion would increase the power allowance.) In this
case the power budget required to fill in the spectrum from the knee
to the ankle is found to be $d\epsilon_{\rm CR}/dt \simeq 2 \times
10^{39}~{\rm erg/s}$~\cite{Gaisser:2006xs}.

We also note that recent data from KASCADE-Grande~\cite{::2013dga}
indicate that at $\sim 30~{\rm PeV}$ the flux of protons is about an
order of magnitude smaller than the all-species CR flux. Taken at face
value, this implies that the fraction of the power budget allocated to
nucleons of energy $E_p$ which do not escape the Galaxy is about $0.1$
of the all-species power.  However, light elements possess higher
magnetic rigidity and are therefore more likely to escape the Galaxy.
From the functional form of $\tau(E/Z)$ above, we estimate the
survival probability for protons at $30~{\rm PeV}$ to be 46\% of that
at $E_{\rm knee}$.  This leads to a value for the proton fraction of
total flux at injection ($\zeta$) of $\zeta = 0.1 / 0.46 = 0.22$.  In
our analysis, we will consider a wide range for $\bar\zeta\equiv\zeta
A_0/A$, with $0.22 \alt \bar\zeta \alt 0.44$ seemingly the most
realistic range.

\begin{figure}[tbp]
\postscript{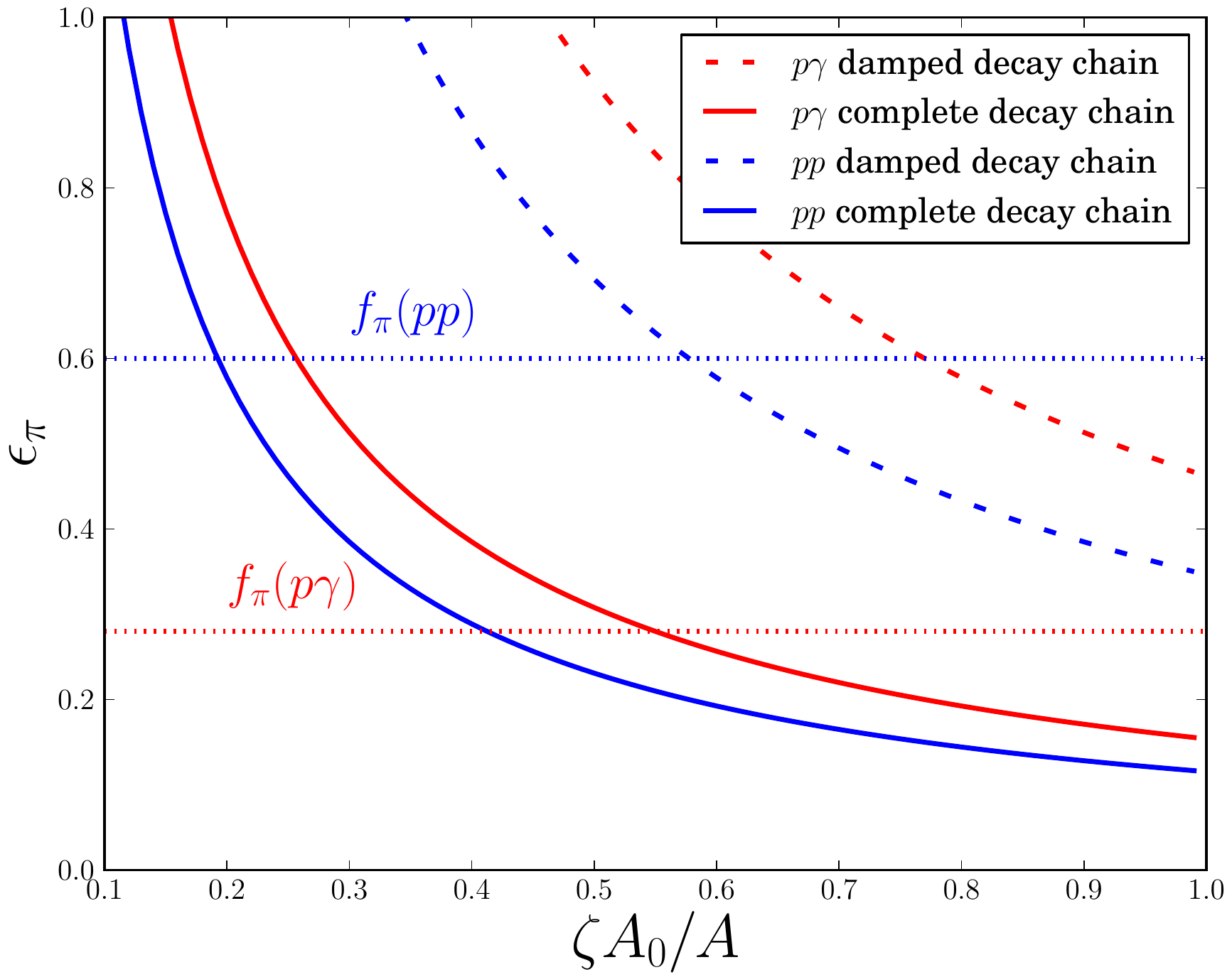}{0.9}
\caption{Total pion energy fractions of parent proton, for favored
  spectral index $\Gamma=2.3$. The average inelasticity
  $f_\pi$ for $pp$ and $p\gamma$ collisions is also shown for comparison.}
\label{fig:epsilonpi}
\end{figure}

Then, we find for $C_{\rm CR}^p$ the particular result
\beq{Ccr2.3}
C_{\rm CR}^p(2.3) = \frac{0.3\times (0.1\,{\rm PeV})^{0.3} \times 2 \bar\zeta \times 10^{39}{\rm erg/s}}{4\pi(10\,{\rm kpc})^2}\,.
\eeq
Finally, inserting Eqs.~\rf{Cnu2.3} and \rf{Ccr2.3} into \rf{epspi}, we get
\beq{epspi2.3} \epspi (2.3) = \left(\frac{1}{16} \right)^{-0.3}
\frac{C_\nu(2.3)}{\epsilon_\nu\, C_{\rm CR}^p(2.3)} =
\frac{0.055}{\bar\zeta \, \epsilon_\nu}\,, \eeq where in the final
expression, we have set $\Gamma$ equal to our favored value of
2.3~\cite{note1}.  If neutrinos are produced in $pp$ collisions, one
can interpret $\epspi$ in terms of the efficiency of transferring
proton energy to all three pion species, $\epsilon_\pi$, by simply
scaling $\epspi$ by $3/2$. Alternatively, if neutrinos are produced in
$p\gamma$ collisions, we scale $\epspi$ by 2~\cite{note2}.  We show
$\epsilon_\pi (\bar\zeta=\zeta A_0/A)$ for all four cases in
Fig.~\ref{fig:epsilonpi}.

In $pp$ collisions, hadronic models predict that $f_\pi \sim 0.6$ of
the ``beam'' proton energy is channeled into
pions~\cite{Frichter:1997wh}.  Since the value of $\epsilon_\pi$
reflects both the inelasticity as well as the fraction of protons
which escape the source without producing pions, we expect
$\epsilon_\pi$ to be smaller than $f_\pi$.  This turns out to be the
case for a complete pion decay chain if $\zeta A_0/A> 0.19$.  Note,
however, that the incomplete pion decay chain requires a considerably
larger fraction, $\zeta A_0/A> 0.59$, which pushes the realm of
plausibility. For $p\gamma$ interactions, $f_\pi \sim
0.28$~\cite{Stecker:1968uc}, thereby excluding the incomplete decay
chain hypothesis for this case.  On the other hand, the complete decay
chain appears to be allowed only for $\zeta A_0/A> 0.56$.

\section{Consistency with Photon Limits and Arrival
  Direction Distribution}
\label{s-cinco}

It is interesting to employ existing limits on high energy photons to
check the plausibility of our hypothesis that the IceCube excess is of
Galactic origin.  $\gamma$ rays are produced by $\pi^0$ decays at the
same optically thin sources where neutrinos are produced by
$\pi^{\pm}$ decay. As described in~\cite{Anchordoqui:2013dnh}, once
can predict a differential $\gamma$ ray flux based the best-fit single
power law $\nu$ flux discussed in this paper, and compare to
measurements.  The CASA-MIA 90\% C.L. upper limits on the integral
diffuse $\gamma$ ray flux, $I_\gamma$ for energy bins,
\begin{equation}
\frac{E_\gamma^{\rm min}}{{\rm GeV}} = 3.30 \times 10^5,\ 7.75 \times 10^5,\ 2.450 \times 10^6 \,,
\label{tres}
\end{equation}
are
\begin{eqnarray}
\frac{I_\gamma}{{\rm cm}^{-2} \ {\rm s}^{-1} \ {\rm  sr}^{-1}} & < & 1.0 \times 10^{-13}, \, 2.6 \times 10^{-14}, \nonumber \\
 & &  2.1 \times 10^{-15}\,,
\label{cuatro}
\end{eqnarray}
respectively~\cite{Chantell:1997gs}. Under the simplifying assumption that there is no
photon absorption, the integral photon fluxes we predict based on our single power law hypothesis
(in units of photons ${\rm cm}^{-2} \ {\rm s}^{-1} \ {\rm sr}^{-1}$), above the
energies specified in (\ref{tres}), are
\begin{eqnarray}
\int_{E_\gamma^{\rm min}} \frac{dF_\gamma}{d\Omega dA dt dE_\gamma} dE_\gamma & = & 4.2 \times 10^{-14}, \ 1.4 \times 10^{-14}, \nonumber \\
& &  3.1 \times 10^{-15} \, .
\label{cinco}
\end{eqnarray}
For the first two energy bins, the predicted fluxes are below the
90\%~C.L. measurements of CASA-MIA, while the last bin slightly exceeds the 90\%
C.L. bound. This does not, however, imply that the Galactic origin hypothesis is
ruled out at 90\% C.L.  First of all, one must keep in mind that sources which
are optically thin up to $E_\gamma \sim 100$~TeV may no be optically thin at
higher energies, suggesting that the importance of photon bounds in
establishing the origin of the IceCube excess should be considered with some
caution~\cite{Razzaque}.  Even if we ignore this caveat, we still do not know the maximum
neutrino energy reached at acceleration sites, so the maximum photon energy is
likewise unknown. In addition, absorption becomes important in the energy regime
covered by the last bin, as mean free path of PeV photons in the CMB is about 10
kpc.

Note that $R_{\rm G} \sim 10~{\rm kpc}$,  leading to an interesting
signature: Photons coming from ``our half'' of the Galaxy will be largely
unattenuated, while those from the farther half will be significantly
attenuated.  Since both photons and neutrinos point back to the sources,
coordinated comparisons of neutrino and photon data will facilitate a completely
new exploration of the highest-energy Galactic sources.  As described
in~\cite{Anchordoqui:2013dnh}, taking into account absorption of the photon flux
for $E_\gamma^{\rm min} > 1$~PeV leads to about a 12\% reduction in the
predicted photon flux.  Furthermore, varying the photon maximum energy cutoff of
Eq.~(\ref{cinco}) to,
\begin{equation}
\frac{E_\gamma^{\rm max}}{\rm PeV} = 6, \ 7, \ 8  \,,
\end{equation}
we obtain
\begin{eqnarray}
\int_{E_\gamma^{\rm min} }^{E_\gamma^{\rm max}} \frac{dF_\gamma}{d\Omega dA dt dE_\gamma} dE_\gamma  & =  & 2.1 \times 10^{-15},\ 2.3 \times 10^{-15}, \nonumber \\
& & 2.4 \times 10^{-15} \, .
\end{eqnarray}
From the discussion above, we can see there are several ways to comply with the
CASA-MIA bound.  For instance, $E_\gamma^{\rm max} = 6~{\rm PeV}$ is already
consistent with the measured bound, even without absorption.  For higher
energies, absorption provides enough reduction of the photon flux to retain
consistency with measurements.  It is also worth noting that the comparison
discussed here is based on experimental bounds on the all-sky $\gamma$ ray flux.
A more rigorous comparison would involve measurements on the diffuse $\gamma$
fay flux within about $15^\circ$ of the Galactic plane. The CASA-MIA Collaboration has
in fact studied $\gamma$ ray emission from the direction of the Galactic plane,
reporting the flux limits as a fraction of the CR flux~\cite{Borione:1997fy}
rather than an integral bound.  Comparing the relative fractions from the
all-sky analysis to the Galactic plane analysis indicates that constraining the
observation to the Galactic plane region does indeed lead to tighter
constraints; the first two energy bins discussed above are roughly saturated at
the 90\% C.L. while the bound for the highest energy bin remains roughly the
same. Only the IceCube collaboration has thus far reported constraints between 1
and 10 PeV.  Bounds from the IceCube 40 string
configuration~\cite{Aartsen:2012gka}, are not restrictive enough to challenge
the Galactic origin hypothesis.  However, within 5 years of data taking with the
complete IceCube configuration of 86 strings, enough statistics will be gathered
to elucidate the $\nu - \gamma$ ray connections.

Finally we comment on the consistency between the arrival direction
distribution of the IceCube excess and the hypothesis that the sources
are nearby.  Fourteen of the 26 reported neutrino events arrive from
within about $15^{\circ}$ of the Galactic plane, including one of the
two highest energy events, which coincides with the Galactic center
(within errors).  The highest energy event is outside of this angular
window, but (as noted in~\cite{Ahlers:2013xia}) does correspond with a
possible hotspot in the IceCube photon search~\cite{Aartsen:2012gka}.
This could reflect emission of neutrinos and $\gamma$ rays from a
common, nearby source, as $\gamma$ rays do not survive propagation
further than $\sim 10$~kpc. The recently discovered large reservoir of
ionized gas extending over a large region around the Milky
Way~\cite{Gupta:2012rh} could provide the target material required for
neutrino production outside the Galactic disk in models in which
proton diffusion extends to the Galactiuc
halo~\cite{Jones:2000qd}. However, given the current statistics and
the insufficient understanding of the atmospheric (in particular the
prompt neutrino~\cite{Lipari:2013taa}) background, the arrival
direction distribution neither favors nor disfavors a Galactic
origin~\cite{Ahlers:2013xia,Neronov:2013lza}.  More data are required
to settle the issue.

\section{Conclusions}
\label{s-seis}

Summarizing, we embrace this joyous moment that appears to be the dawn
of neutrino astronomy, by investigating the hypothesis of a single
power-law Galactic neutrino flux, and investigating several further 
consequences of the hypothesis. Implicit in our phenomenological
analysis is the assumption that there exist Galactic cosmic ray
sources which are both optically thin and capable of generating
protons with energies well beyond the knee feature, and neutrinos with
energies around $E_\nu = E_p / 16$.  As discussed above, this
assumption may stress acceleration models, but is not excluded by
current cosmic ray observations. Combining the assumption that sources
with these requisite conditions exist with the hypothesis that the
observed neutrino spectrum can be characterized by a single power law
leads to three interesting ramifications.

We find that a spectral index of
$\sim 2.3$ is consistent with the data over the range 50~TeV-10~PeV,
at 1.5$\sigma$.  A shallower spectrum overproduces events in the null
region above $\sim 2$~PeV, while a steeper spectrum fails to match the
event rate below a PeV to that at 1-2~PeV.  The first ramification is
that we identified a discriminator between the ``dip model'' for the
Galactic to extragalactic crossover, and the ``ankle model.''  The
discriminator is the termination energy of the neutrino spectrum.  If
it is 1-2~PeV, then the ``dip model'' is favored; if it is 8-10~PeV,
then the ``ankle model'' is favored.  Secondly, we identified a means
of discriminating between competing models for explaining the knee
feature.  If the knee results from an overlay of spectra for two types
of sources, one of which is reaching its acceleration endpoint, we
expect to see a break in the neutrino spectrum around 190~TeV.  If the
knee is a consequence of rigidity dependent leakage from the Galaxy,
we expect no such break in the neutrino spectrum.  The third
ramification is that although the resulting energy fraction
transferred from parent proton to daughter pions is only 2 to 3 times
below the Waxman-Bahcall (WB) bound~\cite{Waxman:1998yy}, the neutrino
flux beyond 1 PeV requires a steep spectrum $\propto E_\nu^{-2.3}$.
This has the unfortunate consequence of requiring 1 order of magnitude
more years, or 1 order of magnitude larger detector volume, to produce
the same event numbers hoped for from saturation of the original WB
bound beyond 1~PeV.

Thus far the IceCube excess is consistent with a
Galactic origin, so we have included all data in our analysis.  In the
future, however, the data may well show evidence of extragalactic
sources.  In this case the analysis presented here can be repeated
with cuts to exclude extragalactic "contamination," {\it e.g.}, by
requiring events to arrive from within $15^\circ$ or so of the
Galactic plane.

\acknowledgments{We thank Francis Halzen, Ruoyu Liu, and Soeb Razzaque
  for discussion.  This work was supported by the US NSF grant
  numbers: CAREER PHY1053663 (LAA); PHY-0757959 (HG); PHY-1205854
  (TCP); PHY-1068696, PHY-1125897 (AVO); US DoE grant
  DE-FG05-85ER40226 (TJW); NASA 11-APRA-0058 (LAA, TCP), 11-APRA-0066
  (AVO); and UWM Physics 2013 Summer Research Award (MHL).}


\begin{thebibliography}{99}
\bibitem{note1} Though the damped pion decay chain does not yield
    flavor equipartition on Earth, any deviation from $\Phi_0^{\rm total}$
    falls in the range of uncertainty.
\bibitem{note2}
Resonant $p\gamma$ interactions produce twice as many neutral pions as charged pions. Direct pion production via virtual meson exchange contributes about 20\% to the total cross section, almost exclusively producing $\pi^+$. Hence, $p\gamma$ interactions produce roughly equal numbers of $\pi^+$ and $\pi^0$.
\end{thebibliography}
\end{document}